\documentclass[pdflatex,sn-nature]{sn-jnl}

\usepackage{graphicx}%
\usepackage{multirow}%
\usepackage{amsmath,amssymb,amsfonts}%
\usepackage[title]{appendix}%
\usepackage{xcolor}%
\usepackage{textcomp}%
\usepackage{manyfoot}%
\usepackage{booktabs}%
\usepackage{listings}%

\usepackage{orcidlink} 

\begin{document}

\title[WorkflowHub: a registry for computational workflows]{WorkflowHub: a registry for computational workflows}

\author[1]{
    \fnm{Ove Johan Ragnar} \sur{Gustafsson}~\orcidlink{0000-0002-2977-5032}} \email{johan.gustafsson@unimelb.edu.au}
\affil[1]{
    \orgdiv{Australian BioCommons},
    \orgname{University of Melbourne},
    \orgaddress{
        \city{Melbourne},
        \state{Victoria},
        \country{Australia}
    }
}

\author[2]{
    \fnm{Sean R.} \sur{Wilkinson}~\orcidlink{0000-0002-1443-7479}} \email{wilkinsonsr@ornl.gov}
\affil[2]{
    \orgdiv{Oak Ridge Leadership Computing Facility},
    \orgname{Oak Ridge National Laboratory},
    \orgaddress{
        \city{Oak Ridge},
        \state{Tennessee},
        \country{USA}
    }
}

\author[3]{
    \fnm{Finn} \sur{Bacall}~\orcidlink{0000-0002-0048-3300}} \email{finn.bacall@manchester.ac.uk}
\affil[3]{
    \orgdiv{Department of Computer Science},
    \orgname{University of Manchester},
    \orgaddress{
        \city{Manchester},
        \country{UK}
    }
}

\author[3,4]{
    \fnm{Stian} \sur{Soiland-Reyes}~\orcidlink{0000-0001-9842-9718}} \email{soiland-reyes@manchester.ac.uk}
\affil[4]{
    \orgdiv{Informatics Institute},
    \orgname{University of Amsterdam},
    \orgaddress{
        \city{Amsterdam},
        \country{The Netherlands}
    }
}

\author[5]{
    \fnm{Simone} \sur{Leo}~\orcidlink{0000-0001-8271-5429}} \email{simone.leo@crs4.it}
\affil[5]{
    \orgname{Center for Advanced Studies, Research, and Development in Sardinia (CRS4)},
    \orgaddress{
        \city{Pula},
        \state{Cagliari},
        \country{Italy}
    }
}

\author[5]{
    \fnm{Luca} \sur{Pireddu}~\orcidlink{0000-0002-4663-5613}} \email{luca.pireddu@crs4.it}

\author[3]{
    \fnm{Stuart} \sur{Owen}~\orcidlink{0000-0003-2130-0865}} \email{stuart.owen@manchester.ac.uk}

\author[3]{
    \fnm{Nick} \sur{Juty}~\orcidlink{0000-0002-2036-8350}} \email{nick.juty@manchester.ac.uk}

\author[6,7]{
    \fnm{José Mª} \sur{Fernández}~\orcidlink{0000-0002-4806-5140}} \email{jose.m.fernandez@bsc.es}
\affil[6]{
    \orgname{Barcelona Supercomputing Center (BSC)},
    \orgaddress{
        \country{Spain}
    }
}
\affil[7]{
    \orgname{Spanish National Bioinformatics Institute (INB)},
    \orgaddress{
        \country{Spain}
    }
}

\author[8]{
    \fnm{Tom} \sur{Brown}~\orcidlink{0000-0001-8293-4816}} \email{brown@izw-berlin.de}
\affil[8]{
    \orgname{Leibniz Institute for Zoo- and Wildlife Research},
    \orgaddress{
        \city{Berlin},
        \country{Germany}
    }
}

\author[9,10]{
    \fnm{Hervé} \sur{Ménager}~\orcidlink{0000-0002-7552-1009}} \email{herve.menager@pasteur.fr}
\affil[9]{
    \orgdiv{Institut Pasteur},
    \orgname{Université Paris Cité},
    \orgaddress{
        \city{Paris},
        \country{France}
    }
}
\affil[10]{
    \orgdiv{CNRS, UMS 3601},
    \orgname{Institut Français de Bioinformatique},
    \orgaddress{
        \city{Evry},
        \country{France}
    }
}

\author[11]{
    \fnm{Björn} \sur{Grüning}~\orcidlink{0000-0002-3079-6586}} \email{bjoerngruening@gmail.com}
\affil[11]{
    \orgname{Albert-Ludwigs-Universität Freiburg},
    \orgaddress{
        \city{Freiburg},
        \country{Germany}
    }
}

\author[6,7]{
    \fnm{Salvador} \sur{Capella-Gutierrez}~\orcidlink{0000-0002-0309-604X}} \email{salvador.capella@bsc.es}

\author[12]{
    \fnm{Frederik} \sur{Coppens}~\orcidlink{0000-0001-6565-5145}} \email{frederik.coppens@vib.be}
\affil[12]{
    \orgdiv{VIB Data Core},
    \orgname{VIB Technologies},
    \orgaddress{
        \city{Ghent},
        \country{Belgium}
    }
}

\author*[3]{
    \fnm{Carole} \sur{Goble}~\orcidlink{0000-0003-1219-2137}} \email{carole.goble@manchester.ac.uk}


\abstract{The rising popularity of computational workflows is driven by the need for repetitive and scalable data processing, sharing of processing know-how, and transparent methods. As both combined records of analysis and descriptions of processing steps, workflows should be reproducible, reusable, adaptable, and available. Workflow sharing presents opportunities to reduce unnecessary reinvention, promote reuse, increase access to best practice analyses for non-experts, and increase productivity. In reality, workflows are scattered and difficult to find, in part due to the diversity of available workflow engines and ecosystems, and because workflow sharing is not yet part of research practice. 

 WorkflowHub provides a unified registry for all computational workflows that links to community repositories, and supports both the workflow lifecycle and making workflows findable, accessible, interoperable, and reusable (FAIR). By interoperating with diverse platforms, services, and external registries, WorkflowHub adds value by supporting workflow sharing, explicitly assigning credit, enhancing FAIRness, and promoting workflows as scholarly artefacts. The registry has a global reach, with hundreds of research organisations involved, and more than 700 workflows registered. 
}

\keywords{workflows, registry, FAIR}



\maketitle

\section{Introduction}\label{sec:introduction}
In an era of Big Data and data-driven science, the need for repetitive, scalable, reproducible and quality-assured data processing and analysis methods has contributed to a surge in popularity for computational workflows \cite{roadmap2022}. The past two decades have seen a handful of workflow management systems (WfMS) expand to hundreds \cite{amstutz_existing_nodate}, and workflows applied across a growing number of domains, including biosciences \cite{maier_ready--use_2021}, astronomy \cite{freudling_adaptive_2023} and the physical sciences \cite{nichols_toward_2020}.

In brief, computational workflows are a special kind of software for handling multi-step, multi-code data pipelines, analysis, and simulations, and are intended to automate data-handling processes. They come in many forms, but typically share certain features: a high-level language executed by a dedicated WfMS, which manages data flow and code execution; a composition of modular code or workflow building blocks that can be remixed; and a tendency to be closely associated, even intertwined with the data on which they will operate \cite{gil_examining_2007}. Important scientific goals like repeatability, replicability, and reproducibility become more realistic when scientists specify their experiment’s analysis processes as a computational workflow \cite{cohen-boulakia_scientific_2017}. For example, computational workflows have become central to major international science missions that require systematic, reproducible, and shared data analysis. Recent examples include the global response to the COVID-19 pandemic and the analyses of SARS-CoV-2 \cite{maier_ready--use_2021}, and the large-scale sequencing efforts currently in-flight for the Vertebrate Genomes Project (VGP) \cite{lariviere_scalable_2024}. While these large consortia with defined collaborative research programs are key drivers for computational workflow creation and deployment, workflows are also being adopted across scientific disciplines as their computational requirements increase, and the emphasis on reproducibility and portability increases \cite{the_icgctcga_pan-cancer_analysis_of_whole_genomes_consortium_pan-cancer_2020,reiter_streamlining_2021,wratten_reproducible_2021,patel_reproducibility_2022,goble_eosc-life_2023}.

Increasingly, scientists are also being asked to share their data and associated research objects (i.e. software), in ways others can reuse (e.g. the Nelson Memo\footnote{\url{https://www.whitehouse.gov/wp-content/uploads/2022/08/08-2022-OSTP-Public-Access-Memo.pdf}}, NASA SPD-41a\footnote{\url{https://smd-cms.nasa.gov/wp-content/uploads/2023/08/smd-information-policy-spd-41a.pdf}}). The idea is to accelerate scientific progress and spur innovation by enabling scientists to avoid reinventing each others’ work, and to explicitly support confidence in published results by removing ambiguity surrounding the approach taken to create research outcomes. In addition, scientific activity often includes the exploration of analysis variance; modifying workflows to understand effects and changes on data products is simpler when those workflows are clearly described and available. To this end, Wilkinson et al. published guiding principles for scientific data management and stewardship, providing guidelines for making data and other research objects Findable, Accessible, Interoperable, and Reusable (FAIR) by others \cite{wilkinson_fair_2016}. The FAIR principles have sparked an entire movement in the international community towards adopting FAIR practices, and further work has been undertaken to extend the principles to research software \cite{barker_introducing_2022}, AI models \cite{Huerta2023-sf}, and computational workflows \cite{wilkinson2024}. A fundamental step towards supporting FAIR workflows \cite{goble_fair_2020} is to enable the sharing of workflows and their descriptions \cite{gil2009} and make them findable.

Researchers find software by searching: the web (i.e. search engines), public software project repositories (e.g. GitHub), the literature, mailing lists, discussion groups (e.g. StackOverflow), dependencies in the software itself, relevant registries (e.g. CRAN\footnote{\url{https://cran.r-project.org/}}, communities of practice like nf-core \cite{ewels_nf-core_2020,di_tommaso_nextflow_2017}), and even social media \cite{stevens_understanding_2022}. High-quality machine-processable metadata markup is needed to make workflows more findable and understandable in such a ``search context'': in other words, the descriptors for workflows must be themselves standardised, accessible, and discoverable. Current mechanisms for sharing workflows do not achieve this outcome for the entire workflow ecosystem. Sharing source code includes options such as version control platforms (e.g. GitHub\footnote{\url{https://github.com/}} or GitLab\footnote{\url{https://about.gitlab.com/}}), and WfMS-specific curated git repositories (e.g. Intergalactic Workflow Commission (IWC)\footnote{\url{https://github.com/galaxyproject/iwc}} \cite{the_galaxy_community_galaxy_2024}, nf-core, Snakemake catalogue\footnote{\url{https://snakemake.github.io/snakemake-workflow-catalog/}} \cite{molder_sustainable_2021}). Creators can also publish their workflows, either in public generalist repositories (e.g. Zenodo\footnote{\url{https://zenodo.org/}}, DataVerse\footnote{\url{https://dataverse.org/}}), conventional journals (e.g. GigaScience\footnote{\url{https://academic.oup.com/gigascience}}) or software journals (e.g. the Journal of Open Source Software, JOSS\footnote{\url{https://joss.theoj.org/}}). Finally, a creator can register the workflow using either a platform-specific (e.g. Knime Community Hub\footnote{\url{https://hub.knime.com/}}, BinderHub\footnote{\url{https://binderhub.readthedocs.io/en/latest/}}, nf-core\footnote{\url{https://nf-co.re/}}) or platform-agnostic solution. The latter includes Dockstore, a registry that supports the sharing and running of containerised tools, workflows and notebooks across diverse cloud computing environments\footnote{\url{https://dockstore.org/}} \cite{yuen_dockstore_2021}.

While critical for software, discovery of these many resources and platforms can be impeded by non-standardised descriptors that are not necessarily visible to search. Even once a workflow has been found, a divergent ecosystem does not lend itself to better integration of services, adoption of standards, or achieving FAIR outcomes for workflows. A registry that serves as a hub for these various mechanisms, and their specific benefits, would begin to address these challenges. It could support workflow developers to share and gain credit for their work, integrate with the platforms, services and infrastructures that developers and users rely on to both create and use workflows, and support making workflows FAIR. Structurally, a registry should be flexible, extensible, and use internationally recognised standards that accommodate rich metadata. To capture and present the breadth of the global computational workflow ecosystem back to the research community, a registry should be agnostic to domains and WfMS, and embrace community standards. Finally, it should provide mechanisms that can link workflows to other digital objects that provide context for a research project, including documents, standard operating procedures (SOPs) and publications.

Here, we present a public and inclusive registry dedicated specifically to the sharing of computational workflows: WorkflowHub\footnote{\url{https://workflowhub.org}} \cite{goble_implementing_2021,goble_workflowhub_2022,goble_eosc-life_2023}. WorkflowHub is designed to allow any scientist, regardless of expertise level, to contribute and share computational workflows. It indexes workflows from any scientific domain, in any format, in any workflow language, regardless of whether it uses a WfMS. Here, we describe in detail how WorkflowHub's structure, design, standards, community engagement, and continued evolution support: 1) collaboration, sharing and credit for workflow developers, projects, and consortia; 2) integration with added-value services, platforms, and capabilities that support the workflow life cycle (i.e. creation, version control, execution, maintenance, reuse and citation); and 3) wizards and inbuilt features that ease the process of sharing workflows alongside the constellation of associated digital artefacts that give a workflow its scientific context.

\section{Results}\label{sec:results}

\subsection{\textbf{A registry for computational workflows}}

\begin{figure}[htbp]
    \centering
    \includegraphics[width=0.9\textwidth]{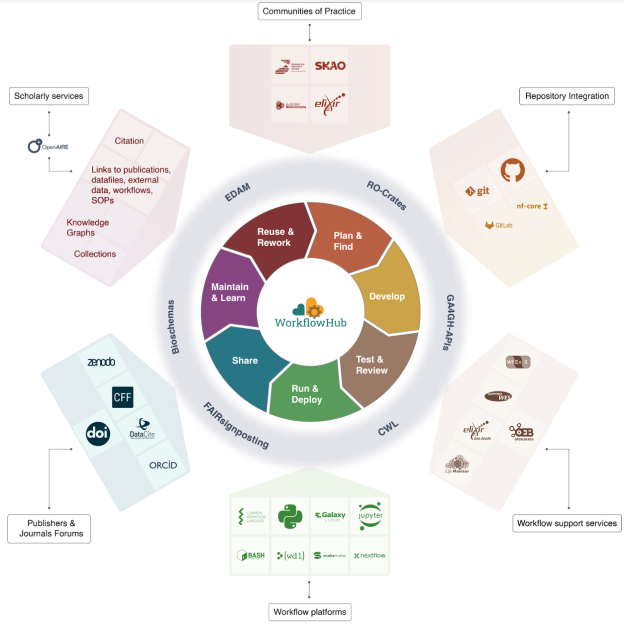}
    \caption{WorkflowHub connects to platforms, services, and resources that support a workflow's life cycle \cite{courbebaisse_research_2023}. A researcher initially needs to \textbf{Plan \& Find}, where they either plan for a particular analysis and find existing workflows (i.e. using a registry), or \textbf{Develop} a new workflow. WorkflowHub integrates with Git repositories (e.g. GitHub, GitLab), and Git-supported communities (e.g. nf-core), to support development. A workflow requires \textbf{Test \& Review} to \textbf{Run \& Deploy}, and here WorkflowHub connects to support services (e.g. LifeMonitor, bio.tools, Sapporo WES, WfExS) and welcomes diverse workflow platforms that aid deployment (e.g. CWL, Snakemake, Galaxy, Jupyter, Python, BASH, WDL, Nextflow). A creator needs to \textbf{Share} a workflow and can benefit from WorkflowHub's use of citation infrastructures and standards (i.e CITATION.cff, Zenodo, DataCite, DOI and ORCID). In the \textbf{Maintain \& Learn} stage, maintenance, and also understanding of a workflow by other researchers, becomes critical as it impacts workflow \textbf{Reuse \& Rework}, where a workflow is either reused, or adapted, by other researchers to suit their requirements. WorkflowHub supports these stages through registration of digital objects that enrich a workflow (e.g. documents, publications, SOPs), the ability to create Collections and workflow citations based on DOIs, and ultimately through the connections created to knowledge graphs. WorkflowHub also enables communities of practice to benefit from all its integrations and connections, ensuring that they can reuse or rework workflows from across the globe. The entire support framework is enabled by the implementation of standards that allow WorkflowHub to interact with the ecosystem and truly act as a ``Hub'': EDAM, Research Object Crates (RO-Crates), GA4GH APIs, abstract Common Workflow Language (CWL), FAIR Signposting, and Bioschemas.}\label{fig:1}
\end{figure}

The WorkflowHub is a registry for describing, sharing and publishing scientific computational workflows, irrespective of their type, development and maintenance location, or discipline. On the landing page for WorkflowHub, a new user is presented with a description of the platform and its purpose, the latest workflow additions, what content is discoverable, and how to join the WorkflowHub community. Underpinning WorkflowHub is the implementation of open tools and standards, which are further described herein. Collaborating Teams are supported by registry features that support workflow reuse, and include integration with native workflow repositories, assignment of credit, import and export, and the creation of curated Collections of workflows that are enriched by other digital objects (e.g. publications, SOPs). Figure~\ref{fig:1} provides a conceptual view of the WorkflowHub’s capabilities and its relationship to the workflow development and publishing ecosystem outlined above and discussed later in more depth.

The registry is agnostic to domain, discipline and workflow type, supporting its adoption by a wide spectrum of researchers and other stakeholders. As a result, at the time of writing (October 2024), the registry already indexes workflows ranging from biodiversity, to astronomy and particle physics (see workflows list on WorkflowHub\footnote{\url{https://workflowhub.eu/workflows}}), with 764 workflows registered for an array of workflow types\footnote{\url{https://workflowhub.eu/workflow_classes}} and 840 registered users from 236 Organisations across 35 countries.

\begin{figure}[htbp]
    \centering
    \includegraphics[width=0.9\textwidth]{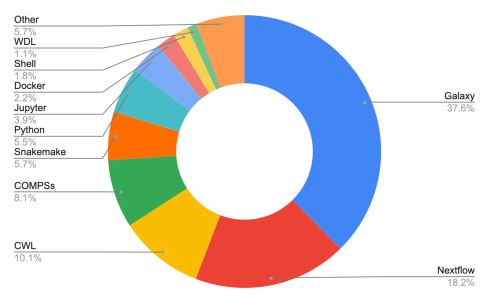}
    \caption{Workflow types registered with WorkflowHub.}\label{fig:2}
\end{figure}

WorkflowHub was launched in 2020, as part of the EOSC-Life Workflow Collaboratory \cite{goble_eosc-life_2023}, to support the registration of workflows required for the response to the COVID pandemic. WorkflowHub now houses 66 COVID-related workflows\footnote{\url{https://workflowhub.eu/search?utf8=\%E2\%9C\%93&q=covid\#workflows}}, including those that support the ongoing global analysis of intra-host variation as new samples become available \cite{maier_ready--use_2021, baker_no_2020}. This outcome demonstrates a central ambition of WorkflowHub: to be of practical use in advancing the application of computational workflows in research science by \textit{supporting the needs of the communities that it serves}.

WorkflowHub meets community requirements and supports the workflow life cycle in three key ways. Firstly, the registry provides structures that directly support collaboration, sharing knowledge and distributing credit. Secondly, woven into this structure are multiple integrations with other elements of the global research ecosystem that support the workflow life cycle: creation, development, discovery, reuse, and citation of workflows. Finally, WorkflowHub provides a registration wizard that guides users in leveraging these structures and integrations. This approach is deliberate and will continue to evolve in lock step with the requirements of the community. In the following sections, we will first describe how the data model and metadata framework of WorkflowHub support the registry's core functions - namely registering, finding, and launching workflows and associated digital data objects. In turn we highlight how the registry design, including the use of wizards to guide best practice, allows it to act as an integrating hub across the workflows ecosystem and to support each stage of the workflow life cycle. Finally, we will describe how the WorkflowHub engages with the workflow community and highlight some key use cases for the registry.

\subsection{\textbf{A data model that reflects the real-world collaborations that create workflows}}

Science is a collaborative enterprise, and infrastructure platforms should reflect this quality to be of practical use in accelerating science missions. Research programs and projects are also intertwined with diverse computational approaches and ways of sharing research outcomes, and these are subject to the same requirements for findability, credit and impact assessment \cite{gil_examining_2007}. As a result, WorkflowHub is structured to reflect real-world collaboration and assign complex credit well. 

\begin{figure}[htbp]
    \centering
    \includegraphics[width=0.9\textwidth]{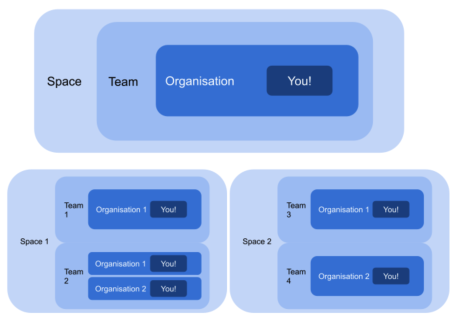}
    \caption{A guide to the structures in WorkflowHub. You, the user, belong to one or more Organisations (i.e. affiliations). You can also belong to one or more Teams, each of which also needs to belong to a single Space \textbf{(top)}. You can nominate which Organisations you wish to use for the different Teams that you have created or joined, and you can belong to multiple Teams in the same Space, as well as multiple Teams in other Spaces \textbf{(bottom)}. Image reused with permission from WorkflowHub documentation.}\label{fig:3}
\end{figure}

The data model of WorkflowHub provides access to three elements for every registered user: Organisations, Teams, and Spaces. A user can specify one or more Organisations (i.e. affiliations) as part of their user profile. They also must belong to at least one Team, which must also belong to an administrative Space. If existing Teams are not suitable or appropriate, a user can create a new Team. A member of a Team can specify Organisations for each Team they join, which allows users to be related to different Organisations for different Teams. Multiple creators and Teams can be specified for a single workflow, additional credit can be assigned to contributors, and a distinction can be made between creators and submitters. 
In effect, users belong to Teams, which belong to Spaces, and in this way credit is able to cascade as required from a workflow to creators, contributors, submitters, the Teams and Spaces to which they belong, the consortia and Organisations that these represent, and even new workflows which are derived from the original (see Figure~\ref{fig:3}). This nested structure is capable of addressing sharing and credit for a diverse set of workflow contributors, including, but not limited to, individuals (e.g. workflow developers), research groups, institutions (e.g. universities), communities of practice (e.g. nf-core), and major research consortia (e.g. Biodiversity Genomics Europe (BGE)\footnote{\url{https://biodiversitygenomics.eu/}, \url{https://doi.org/10.3030/101059492}}). For example, single users or research groups may only require a single Team to represent their workflow(s), and in this case they would add their Team to the default “Independent Teams” Space. However, a consortium representing multiple research groups or institutes may create a distinct Team for each one of its collaborating entities, and add the Teams to either the Independent Teams Space, or create a new Space to administer all these Teams. An individual, group, or consortium can thus establish a presence on WorkflowHub that reflects their real world structure, and which WorkflowHub then uses to assign credit. 

The structure of WorkflowHub can also support an entire community of practice, whereby Spaces and Teams are used to organise contributors, share workflows, and support the sharing of knowledge within that community. This is achieved by registering workflows and linking them to other research outputs, including events, presentations, documents, publications, data files and SOPs - these assets can either be hosted by WorkflowHub or added by reference. A workflow developer can even nominate relevant community channels where they can connect with users, and curated “Collections” of workflows can be constructed to help a community manage specific sets of relevant workflows, and other outputs, that may span many scientific applications, workflow languages and research programs (e.g. Threatened Species Initiative\footnote{\url{https://threatenedspeciesinitiative.com/about-tsi/}} annotation workflows\footnote{\url{https://workflowhub.eu/collections/23}}). To streamline the onboarding process for consortia and larger projects, a generic set up guide \cite{soiland-reyes_guide_2024}, and multiple guides for specific consortia \cite{soiland-reyes_bge_2024, soiland-reyes_biodt_2024, goble_by-covid_2024}, have been created. These guides describe the steps required to get set up on WorkflowHub and use the platform effectively.

\subsection{\textbf{A Hub for workflows}}

As the name of the registry suggests, the underlying aim of WorkflowHub is to act as a “Hub”, and specifically a Hub that helps to connect the workflows ecosystem and ease the interoperation of its constituent platforms and services. To realise this, WorkflowHub relies on a web-friendly metadata framework that simultaneously supports data representation within the registry itself, while also acting as a foundation for exchange of data and metadata between the platforms and services that are described in Figure~\ref{fig:1}. Here, we provide more details for three core areas that underpin this “Hub” functionality \cite{goble_eosc-life_2023}. 

WorkflowHub participates in the EOSC federated Authentication and Authorization Infrastructure through LS-Login and the OAuth2 protocol, in addition to also supporting authentication via GitHub. This feature allows systems interacting with WorkflowHub to identify users as the same individual across different systems – even with their identity provided by their participating institutional accounts – enabling single-sign on and smart authorization decisions on accessing and operating on workflows, their metadata and other resources across the ecosystem forming around WorkflowHub.

To help describe workflows and their components, the registry uses three profiles from bioschemas \cite{gray_bioschemas_2017}: Computational Tool\footnote{\url{https://bioschemas.org/profiles/ComputationalTool/1.0-RELEASE}}, Computational Workflow\footnote{\url{https://bioschemas.org/profiles/ComputationalWorkflow/1.0-RELEASE}}, and Formal Parameter\footnote{\url{https://bioschemas.org/profiles/FormalParameter/1.0-RELEASE}}. Even though these are a part of the bioschemas effort, the profiles support a discipline independent and standardised way of describing workflows and their components. In addition, and owing to close collaboration with the Common Workflow Language (CWL) \cite{crusoe_methods_2022, amstutz_common_2016} community, the CWL workflow specification is encouraged as a workflow language independent way of describing a WorkflowHub entry. This is the so-called “Abstract CWL”. This format can even hold semantic annotations, which WorkflowHub leverages to extract the typing of workflow inputs and outputs, as well as EDAM ontology concepts for Topics and Operations \cite{ison_edam_2013}. 

RO-Crate \cite{soiland-reyes_packaging_2022} is a standard for FAIR Research Objects. It was developed by the community to package a workflow with the components required to understand and execute that workflow. The additional packaged components may include test data, Abstract CWL, diagrams, publications, and SOPs, as well as the flat metadata file that provides the context for all these assets\footnote{\url{https://www.researchobject.org/ro-crate/}} \cite{soiland-reyes_packaging_2022}. The implementation of RO-Crate is central to the ability of WorkflowHub to interoperate and exchange Workflow-RO-Crate\footnote{\url{https://w3id.org/workflowhub/workflow-ro-crate/1.0}} data with the ecosystem in Figure~\ref{fig:1} (e.g. Zenodo archiving) and for exposing workflows as FAIR Digital Objects \cite{soiland_reyes_2024_13225792, de_smedt_fair_2020}. 

WorkflowHub can also interoperate with workflow execution platforms that are part of the ecosystem (see Figure~\ref{fig:1}) through its implementation of the GA4GH Tools Registry Service (TRS) API\footnote{\url{https://ga4gh.github.io/tool-registry-service-schemas/}}\footnote{\url{https://www.ga4gh.org/product/tool-registry-service-trs/}}. This means that a user of a TRS enabled analysis platform, like Galaxy, is able to search for and retrieve workflows, without leaving the platform. It also means that WorkflowHub can send workflows to these platforms, ready for execution. 

Ultimately, the impact of these features is threefold. Firstly, WorkflowHub is able to support machine actionability, as described by the FAIR principles \cite{wilkinson_fair_2016}. This underpins the registry's ability to connect to services and platforms that are in use day-to-day by workflow users and developers. Finally, these same users are able to access this ecosystem ``Hub'' using multiple authentication mechanisms and leverage multiple standards when contributing workflows.

\subsection{\textbf{Using WorkflowHub to register, find and launch workflows}}

The primary purpose of WorkflowHub is to allow researchers to register and share workflows. Existing public workflows can therefore be viewed, downloaded and launched without needing to register with WorkflowHub. By extension, the contribution of open access and publicly accessible workflows is also encouraged. However, workflows may be registered privately, or be embargoed. This functionality supports creators of workflows in cases where they would like to (i.e. a workflow is still being developed), or need to, limit access to a specific group of users. User authentication (i.e. login) is required to register content with WorkflowHub, and enables the registry to assign credit and enable citation. 

To contribute content an individual user needs to:

\begin{enumerate}
    \item \textbf{Register} and indicate the Organisations to which they are affiliated. A user can add the following to their profile: a description, their ORCID\footnote{\url{https://orcid.org/}}, contact details (visible to those in shared Teams and Spaces), as well as knowledge and expertise. More advance configurations are also available via a user profile, including the management of OAuth sessions, authorised applications, API applications, and API tokens.
    \item \textbf{Decide which Space on WorkflowHub to use:} a Space is a user-administered section of WorkflowHub that can be used to manage the Teams required for consortia, institutes, or other large research activities. WorkflowHub administers a single default Space called “Independent Teams”. This Space can be used when users do not need to create and manage many Teams, but simply need to create a single Team. All other Spaces are created upon request and administered by those who requested the Space.
    \item \textbf{Create or join at least one Team:} Teams are one or more people working on a particular research activity involving workflows. Every workflow in WorkflowHub is owned by at least one Team. WorkflowHub users must therefore belong to at least one Team, and this Team must belong to a Space (e.g. Independent Teams). In addition to supporting the correct assignment of credit to workflow developers, contributors, and submitters, the Team also enables its members to further describe the context for their workflow development (i.e. background, project description), and serves to promote their contributions to other WorkflowHub users.
\end{enumerate}

Once these steps are complete, a user has the option to register:

\begin{itemize}
    \item \textbf{Core resources} such as workflows and Collections. As workflows do not exist in a vacuum, Collections allow a WorkflowHub user to bring together workflows with any of their other resources and activities (see below) to create a visible and holistic resource that can support workflow reuse.
    \item \textbf{Other resources}, including publications, documents, data files, and SOPs.
    \item \textbf{Activities}, including presentations and events.
\end{itemize}

Registration of workflows can be carried out by manually uploading a workflow file, importing either a RO-Crate or Git repository, and by submission through a REST API\footnote{\url{https://about.workflowhub.eu/developer/ro-crate-api/}}\footnote{\url{https://about.workflowhub.eu/docs/adding-files/}}. As a registry, WorkflowHub is designed to link to workflows held in their native repositories. However, because manual uploading and storage of files is also supported, it can also act as a repository. RO-Crate is used by WorkflowHub as a fundamental unit that underpins upload, download, import and export. Importantly, a user does not need to know how to create or work with RO-Crates, as the registry automatically builds a crate when a workflow is registered using one of the other available mechanisms. 

To streamline the above processes, each stage is guided by inbuilt wizards and users are prompted to carry out the next step. For example, after user registration with WorkflowHub, a user is prompted to join or create a new Team. After registration of a workflow file or workflow repository, the user is prompted to complete the set of metadata suggested to create a well described and FAIR workflow entry.

\begin{figure}[htbp]
    \centering
    \includegraphics[width=0.9\textwidth]{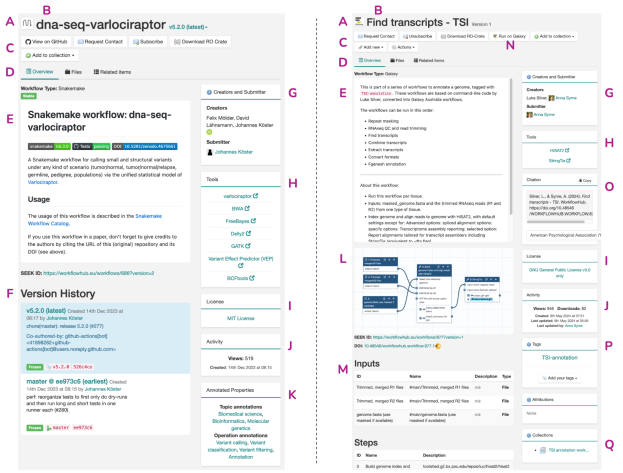}
    \caption{Two example entries in WorkflowHub (left: \cite{koster_snakemake-workflowsdna-seq-varlociraptor_2023}, right: \cite{silver_find_2024}) with sections of the user interface annotated and each entry using the flexible features of WorkflowHub in distinct ways. Entry features include \textbf{A)} workflow type, \textbf{B)} title, \textbf{C)} access panel with links to the source repository (e.g. GitHub), requests to contact the creators, subscribe / unsubscribe, download research object crate (RO-crate), add to a Collection, and in the right hand example access to administrative menus such as Add new (e.g. document, SOP) and Actions (e.g. edit or manage the workflow, including versions and minting DOIs), \textbf{D)} tabs for navigation between the entry overview, the list of files in the entry, and lists of items related to the workflow, including people, Teams, Spaces, Organisations, and other digital objects (e.g. publications, documents, SOPs, other workflows), \textbf{E)} description, which can be imported from Git, if available, \textbf{F)} version history, including Git commits, if available, \textbf{G)} creator and submitter information, \textbf{H)} links to more information about tools that comprise the workflow (i.e. bio.tools registry entries), \textbf{I)} license information, \textbf{J)} activity metrics (i.e. downloads and views), \textbf{K)} ontology concept annotations (e.g. EDAM in the left example entry), \textbf{L)} workflow diagram, \textbf{M)} parsed workflow inputs, outputs and steps for specific WfMS (e.g. Galaxy in the right example entry), \textbf{N)} buttons for launching workflows on execution platforms (e.g. Galaxy for right example entry), \textbf{O)} citation for the workflow (i.e. either using information from a minted DOI or a custom citation (e.g. workflow publication), \textbf{P)} custom tags, and \textbf{Q)} Collections that include the current workflow entry.}\label{fig:4}
\end{figure}

Figure~\ref{fig:4} illustrates two example WorkflowHub entries: the \textit{dna-seq-varlociraptor} Snakemake workflow\footnote{\url{https://workflowhub.eu/workflows/686}} \cite{koster_snakemake-workflowsdna-seq-varlociraptor_2023}, and the \textit{Find transcripts - TSI} Galaxy workflow\footnote{\url{https://workflowhub.eu/workflows/877}} \cite{silver_find_2024}. The top of the entry emphasises the workflow name (Figure~\ref{fig:4}.B) and workflow management system (Figure~\ref{fig:4}.A), along with quick links to the development repository, requesting contact with the authors, subscribing to notifications about changes to the workflow, downloading a workflow RO-Crate, and adding the workflow to a collection. In addition, the workflow creator also has access to administrative options for the workflow entry in this section, which include adding new documents or presentations connected to the workflow, and workflow actions such as adding new versions, requesting a DOI for a specific workflow version, editing the workflow metadata, managing workflow contributors and visibility, and deleting the workflow entry (Figure~\ref{fig:4}.C). The main panel for the entry has three tabs (Figure~\ref{fig:4}.D) that provide a workflow overview (including descriptions, version information, metadata, critical annotations, and activity analytics), access and view capability for the files that were registered, and links to related items (e.g. people, Spaces, Teams, Collections, other workflows).

The examples use many of the key features of WorkflowHub, including those enabled by the WorkflowHub registration wizard, which prompts inclusion of critical metadata: these include title (Figure~\ref{fig:4}.B), workflow management system (Figure~\ref{fig:4}.A), creators (Figure~\ref{fig:4}.G), component tools (Figure~\ref{fig:4}.H), license information (Figure~\ref{fig:4}.I) and ontology annotations (Figure~\ref{fig:4}.K). In addition, the \textit{dna-seq-varlociraptor} workflow made use of the WorkflowHub Git integration to ingest the repository README file (complete with badges), as well as providing the link to the development repository (Figure~\ref{fig:4}.C), the ability to access and view the repository file list natively in WorkflowHub (Figure~\ref{fig:4}.D), and an annotated version history (including commit IDs, Figure~\ref{fig:4}.F). 

To find these workflows within the registry, a user can start by applying a text-based search, or by visiting the complete workflow listing where they can filter by type (e.g. Galaxy, Nextflow), tools used (i.e. using bio.tools identifiers \cite{ison_biotools_2019}, creators, Organisations, Teams, Spaces, and more. Researchers can also refine their searches by making use of faceted browsing and filtering on tags and other annotations, and are able to sort by titles, dates, views and downloads. 

Galaxy's integration with WorkflowHub leverages the GA4GH TRS API, enabling seamless workflow exchange. Researchers can discover, import, and run workflows from WorkflowHub directly within Galaxy. From the WorkflowHub interface users can utilise the ``Run in Galaxy'' button (see Figure~\ref{fig:4}.N), which redirects them to a Galaxy instance and a workflow run form. This interoperability facilitates immediate application and further development of workflows. The use of RO-Crate specification ensures that workflow metadata and components remain accessible and interoperable, aligning with FAIR principles.

\subsection{\textbf{Design that supports the workflow life cycle}}

To support the workflow life cycle \cite{deelman_managing_2006, gil_examining_2007, courbebaisse_research_2023}, WorkflowHub integrates with services that workflow creators use for development, execution, maintenance, testing, citation, and ultimately archiving. These integrations initially form part of the workflow registration wizard, which guides a workflow creator through the process of registering their workflow for the first time. However, they are also accessible during a workflow’s maintenance phase, when the workflow may be updated to modify, improve, or repair its function.

\subsubsection{\textbf{Ease of access}}

A user of WorkflowHub begins by accessing the service via LS-Login. This enables researchers to use credentials from their specific institutions or even other identity-providing platforms (e.g. Google, Apple, ORCID). Authentication via GitHub or a local WorkflowHub account is also supported. This feature also enables WorkflowHub administrators to manage user access rights, and to create a custom combination of access levels that are suitable for specific user groups (e.g. research groups, consortia, international projects). For example, a contributor may simply want to register a single workflow to make it findable and the contributor may be the only user that requires edit access to the workflow, or to the Team to which the workflow belongs, and access rights can be set accordingly. Conversely, a community of practice (e.g. nf-core), or consortium (e.g. BGE) may have multiple contributors, from multiple institutes, that also belong to multiple WorkflowHub Teams. In this case, granular permissions for edit rights can be set at the workflow, Team and Space levels.

\subsubsection{\textbf{Development and versioning }}

Integration with the Git version control system is a key aspect of WorkflowHub that supports workflow creation, development and maintenance. If the workflow registration wizard is provided with a Git repository URL, WorkflowHub will automatically import and parse its metadata. In this case, a workflow creator only needs to review, and potentially update, metadata fields prior to completing the registration process. WorkflowHub’s Git integration supports workflows to remain in their native creation and development environment (i.e. a version control system), avoiding any impact of registration on the workflow’s development and management process. Moreover, it allows for automation of tasks like updating workflow entries in the registry when workflows are versioned (e.g., by using the LifeMonitor GitHub app\footnote{\url{https://lifemonitor.eu/lm_wft_best_practices_github_app}} \cite{goble_eosc-life_2023}). For use cases like Galaxy, where workflows can be created via graphical user interface, rather than scripting, WorkflowHub provides the option to manually upload a workflow file, and step through the wizard manually to enter metadata.

\subsubsection{\textbf{WorkflowHub welcomes all workflows}}

Workflows come in all shapes and sizes, and may even be composed of multiple subworkflows. They may start off as a set of scripts and evolve into a heavily standardised and portable workflow \cite{wratten_reproducible_2021}. They may use one of the many known WfMS \cite{amstutz_existing_nodate}. And, of course, workflows can span virtually every field in the sciences and beyond. In short, the workflow and WfMS ecosystems are diverse. One role of WorkflowHub is to make these ecosystems transparent, and it does this by being agnostic to workflow language, maturity, source, structure, and even scientific quality. Contributions of every workflow type are encouraged\footnote{\url{https://workflowhub.eu/workflow_classes}}. Workflows at any development stage (i.e. work-in-progress) are encouraged, not just those workflows considered to be completed and stable. As such, an indication of the maturity of a workflow can be assigned by the creators, and naturally, this metadata is presented to potential workflow consumers in the registry entry to readily identify workflows that may not be ready for reuse. 

\subsubsection{\textbf{Annotating workflow purpose (i.e. adding metadata)}}

The WorkflowHub registration wizard guides users in provision of metadata. Although Bioschemas metadata profiles are used, the only mandatory metadata fields are the workflow Title and the contributing Team(s). In addition to describing the workflow itself, the metadata wizard can be used to associate a workflow to relevant other workflows, presentations, publications, documents, SOPs and data files.

Two key metadata integrations are in place that allow users to search for and add standard identifiers when editing workflow metadata. This functionality is available for bio.tools software identifiers\footnote{\url{https://bio.tools/}} and the EDAM ontology concept identifiers for both Topics (e.g. genomics) and data transformation Operations (e.g. genome assembly) \cite{ison_edam_2013}. A user can therefore annotate their workflow with persistent links to registry metadata about software that the workflow contains, and standardised shorthand terms that describe its application area and function. In the case of Galaxy, the standard structure of the workflow file is used to extract software tool identifiers and map these automatically to bio.tools \cite{lamothe_evaluation_2023}. A user can also build on these integrations by manually including custom tags and keywords.

\subsubsection{\textbf{Discovery and understanding}}

When a workflow is ready to be shared and reused, a creator can update their workflow maturity from “work-in-progress” to “stable” in the WorkflowHub entry metadata. It is at this point that a workflow needs to be discoverable in multiple ways, and remain accessible in its original location. The reason for this is that not all researchers will necessarily seek to discover workflows in the same way. As a result, WorkflowHub is flexible in its approach. As mentioned earlier, within the registry itself a user can start by applying direct search and filtering to find workflows. External to the registry, the machine-readable, standardised format of WorkflowHub (i.e. Bioschemas) increases the search engine visibility of the rich metadata in a workflow entry. You can interactively explore the impact of workflow registration using evaluator tools such as FAIR-checker \cite{rosnet_fair-checker_2024} and FAIRsoft \cite{del_Pico2022_05_04_490563}.

Finding a workflow is step one for a potential user. Once found, the WorkflowHub entry metadata supports a user to understand the workflow, including its design, content, and purpose \cite{cohen-boulakia_scientific_2017}. For example, “this is a Galaxy type workflow, containing these tools (i.e. links to bio.tools), which are capable of these types of data transformations (i.e. EDAM annotations)”. From a workflow entry that has been annotated with tools, a user can access links to navigate directly to the referenced bio.tools entries to explore and  further understand the components that comprise the workflow. Users can also view the files in the source Git repository (if the workflow was imported from Git), and with a single click visit the repository. It is even possible to subscribe and be notified of changes to entries, removing the need for constant monitoring. 

\subsubsection{\textbf{Execution / reuse}}

WorkflowHub actively supports and develops integrations with workflow execution platforms and services. A key example is the GA4GH TRS API\footnote{\url{https://github.com/ga4gh/tool-registry-service-schemas}}. If an execution platform or system adopts the TRS standard, it can search WorkflowHub for suitable workflows, retrieve those workflows, and execute them, without the need to develop custom integrations with the workflow’s native repository. Examples include, Galaxy, Sapporo \cite{suetake_sapporo_2022} (DNA Data Bank of Japan (DDBJ\footnote{\url{https://ddbj.nig.ac.jp/}}), and the Workflow Execution Service (WfExS)\footnote{\url{https://github.com/inab/WfExS-backend}} \cite{fernandez_secured_2022}, all of which implement the TRS API, either as providers or consumers. As execution platforms (e.g. Galaxy) can make use of the TRS, they are also able to provide inbuilt search interfaces that connect to WorkflowHub and support platform users to find and import a specific workflow. The connection of WorkflowHub to the LifeMonitor service\footnote{\url{https://www.lifemonitor.eu/}}, through the LifeMonitor GitHub app, allows workflow function and status to be reported to maintainers and users through regular automated tests driven by continuous integration (CI) based monitoring (e.g. Planemo automated workflow testing using Galaxy \cite{bray_planemo_2023}). In these cases, WorkflowHub will also include a badge that shows if the tests are passing or failing. An embedded link in the badge takes a user to the LifeMonitor page for the workflow, providing information on the reliability of the workflow over time and the timeliness of the workflow maintainers in solving issues as they arise. In addition, the app can automatically suggest changes to the workflow Git repository that will bring its metadata and structure in line with best practices. Finally, to streamline the inclusion of reuse conditions, licensing information is included in the metadata for workflows registered with WorkflowHub, default sharing and license conditions can be specified for Teams, and these rules can be updated by Team administrators.

\subsubsection{\textbf{Attribution and citation}}

It is important to properly attribute contributions to workflows, and this includes provenance of the entire workflow development process. The RO-Crate format used by WorkflowHub allows for provenance tracking of metadata, ensuring that workflow creators are given the credit they deserve, while also adding accountability. Workflows can be linked to each other using WorkflowHub metadata (i.e. the "attribution" metadata can be used to indicate that a workflow is based on another workflow). The Git integrations described above also support the citation standard CITATION.cff\footnote{\url{https://citation-file-format.github.io/}} \cite{druskat_citation_2021}, so that WorkflowHub can import this file and use its contents to populate workflow entry metadata for creators, including their ORCID. This approach simplifies citing a workflow according to the wishes of the workflow developer. Once a workflow is registered and credit is established within the metadata framework of WorkflowHub, the registry can also, at the push of a button, use DataCite\footnote{\url{https://datacite.org/}} to mint persistent digital object identifiers (DOIs) for workflows and to contribute to the DataCite PID Graph\footnote{\url{https://support.datacite.org/docs/datacite-graphql-api-guide}}. This is the first step to ensuring that workflows can be cited effectively, increasing their visibility and potential impact, and supporting inclusion in the scholarly knowledge graph. 

\subsection{\textbf{User engagement and training}}

WorkflowHub engages and supports a broad set of use cases, including numerous projects and consortia of global significance. Major projects are supported so that their members are able to use WorkflowHub in a way that aligns with the expectations of their project and its funders. Workflow communities (e.g. nf-core, CWL, Snakemake) are directly engaged by the WorkflowHub team, and supported to make their best practice workflows available in the registry.

The main mechanism through which engagement happens is the fortnightly open format WorkflowHub Club meeting\footnote{\url{https://about.workflowhub.eu/\#community}}, where anyone can join the conversation, learn more about the registry, ask questions, and even contribute to the on-going development and evolution of the registry. There are also indirect ways through which workflow creators and users interact with WorkflowHub and its resources. The WorkflowHub Club team creates and presents content for the registry at conferences, in webinars and workshops, and as part of Ask-Me-Anything forum events\footnote{\url{https://about.workflowhub.eu/project/outreach/}}. Registered users of WorkflowHub are also able to ask questions and provide direct feedback via the registry interface. These communications are sent directly to the administrators of WorkflowHub for review and response. Finally, WorkflowHub operates a documentation site where users can access information on how to use the registry\footnote{\url{https://about.workflowhub.eu/docs/}}.

Clear and practical guidelines are required to support users of WorkflowHub. For example, some annotations are consistently missing from workflow entries, and useful features are sometimes overlooked (e.g. Git integration, parsing of CITATION.cff files, linking data to workflows, and linking workflows to each other). This highlights that the rich feature set of WorkflowHub is not necessarily immediately clear, and that guidance in leveraging these features is absolutely critical to support users in achieving best practice. It is important that the provided guidelines also extend to recommendations for how to organise a workflow development repository (i.e. Git repositories). This will enable WorkflowHub to extend the process currently in place for Galaxy IWC to more community repositories: integrating with these repositories in a more standard way to automatically manage the update of workflow versions in WorkflowHub, such that they are in sync with Git releases. As a result, and as highlighted earlier, a general onboarding and set up guide for projects and consortia has been developed \cite{soiland-reyes_guide_2024}, as have multiple consortia specific guides \cite{soiland-reyes_bge_2024, soiland-reyes_biodt_2024, goble_by-covid_2024}. Workflow resources have also been developed for the Galaxy Training Network (GTN \cite{hiltemann_galaxy_2023}) Smörgåsbord events\footnote{\url{https://gallantries.github.io/video-library/modules/ro-crate}} and specific GTN tutorial sets\footnote{\url{https://training.galaxyproject.org/training-material/topics/fair/}}.

Finally, the WorkflowHub team actively identifies opportunities to engage with peer infrastructures to grow the user base of the registry, investigate and create integrations that are of enduring value, and further improve the function of the registry. For example, WorkflowHub is actively fostering a conversation with publishers and journals focused on how to make workflows citable objects in the literature \cite{goble_workflowhub_2024}. WorkflowHub is also being further developed to fully align with the best practice guidelines of the SciCodes Consortium \cite{garijo_nine_2022}, implement the FAIR Principles for Research Software (FAIR4RS) \cite{barker_introducing_2022} and contribute to the CodeMeta specification for software \cite{jones_codemeta_2024}. As described earlier, a registry like WorkflowHub enables the creation of workflows that follow the FAIR principles, from the perspective of data \cite{wilkinson_fair_2016}, software \cite{barker_introducing_2022}, and the unique features of workflows (e.g. abstraction, composition). WorkflowHub is central to discussions in the FAIR Computational Workflows working group for the Workflows Community Initiative (WCI\footnote{\url{https://workflows.community/groups/fair/}}). This working group engages broadly across the global workflows ecosystem (i.e. workflow developers, communities, platforms and services) to develop FAIR principles for workflows \cite{wilkinson2024}. The ultimate aim is to use the outcomes of these engagements to guide the evolution of WorkflowHub as a FAIR registry for workflows.

\subsection{\textbf{Use Cases}}

To effectively support the sharing of workflows, WorkflowHub supports collaborations and communities of practice within the sciences. WorkflowHub contributions span domains such as cancer, COVID, genomics, rare diseases, geosciences, climate, physics, and more. WorkflowHub users span the globe and 35 countries are represented in the registered user list\footnote{\url{https://workflowhub.eu/people}}.

\subsubsection{\textbf{Research consortia \& infrastructures}}

WorkflowHub is an integral platform for consortia and projects. Here we provide details for three specific use cases, EOSC-Life, BGE, and Australian BioCommons.

EOSC-Life was a key use case driver, as it supported the implementation of FAIR computational workflows in the EU by seeking to develop a cloud-based Workflow Collaboratory \cite{goble_implementing_2021} that ultimately resulted in the creation of WorkflowHub. The aim was to create a platform that would support community collaboration on the development, use, and reuse of FAIR computational workflows \cite{goble_fair_2020}, and to do so in a way that bridges research domains and infrastructures \cite{goble_implementing_2021, goble_eosc-life_2023}. WorkflowHub accommodates the diversity of EOSC-Life and ensures the visibility of workflows applied across its many established research infrastructures as they are created and registered \cite{goble_implementing_2021}.

The BGE project is a coming together of two communities of researchers with a common goal of cataloguing biodiversity through genomic resources: the European Reference Genome Atlas \cite{mazzoni_biodiversity_2023} and the European node of the International Barcode of Life consortium (iBOL Europe\footnote{\url{https://iboleurope.org/}}). Providing reference-quality genomes to the community (ERGA) and monitoring biodiversity through DNA barcoding (iBOL Europe) requires the management and processing of vast amounts of data, in an accessible and distributed fashion, relying on input from multiple individuals and institutes. The combination of BGE WorkflowHub Spaces\footnote{\url{https://workflowhub.eu/programmes/25}}, Teams\footnote{\url{https://workflowhub.eu/projects/163}} and Collections\footnote{\url{https://workflowhub.eu/collections/10}} allows individuals to contribute as needed to the projects across the consortium. As workflows have been collected and curated by the community, they in effect also come with a “seal of approval” for external users that wish to replicate the work of ERGA or iBOL Europe. All together, such a structure of publishing and maintaining workflows facilitates BGE in achieving their ambitious goals of cataloguing biodiversity in Europe and bringing together researchers from the biodiversity genomics community.

Australian BioCommons\footnote{\url{https://www.biocommons.org.au/}} is a national infrastructure project that actively supports life science research communities with community scale digital infrastructure \cite{francis_australian_2024}. Rather than building anew, BioCommons aims to adopt fit-for-purpose international platforms and services that, in the case of workflows, can assist with the provision of sophisticated software, analysis capabilities, and digital asset stewardship. WorkflowHub is the primary workflow registry for BioCommons\footnote{\url{https://workflowhub.eu/programmes/8}}, and it is the focal point for sharing the collaborative workflow efforts of BioCommons and its infrastructure partners together with Australian life science researchers\footnote{\url{https://workflowhub.eu/collections/6}}.

\subsubsection{\textbf{Workflow management systems}}

WorkflowHub accepts all workflow types, and includes Galaxy \cite{the_galaxy_community_galaxy_2024}, Snakemake \cite{molder_sustainable_2021}, Nextflow \cite{di_tommaso_nextflow_2017}, job schedulers (e.g. PyCOMPS \cite{tejedor_pycompss_2017}, application-specific types like SCIPION \cite{de_la_rosa-trevin_scipion_2016}, notebooks (e.g. Jupyter\footnote{\url{https://jupyter.org/}}), and even scripting languages (e.g. R \cite{r_core_team_r_nodate} and Python\footnote{\url{https://www.python.org/}} \cite{van1991interactively} (see also Figure~\ref{fig:2}). WorkflowHub provides customised support for WfMS that are critical for specific domain communities (e.g. bioinformatics). This is currently the case for Galaxy, CWL and Nextflow, and additional significant and popular WfMS may be supported with relevant features, when appropriate. As an example, the registry development team, together with Galaxy, have created functionalities for WorkflowHub that include 1) semi-automated registration of new and updated Galaxy IWC workflows, 2) integration with LifeMonitor to further support semi-automated registration but also Planemo workflow test monitoring, and 3) automatic mapping of tool identifiers to the bio.tools registry \cite{goble_eosc-life_2023}. As the CWL community has been closely involved in registry development, WorkflowHub also has in-built functions for parsing CWL, and makes use of Abstract CWL. Most recently, engagement between nf-core and the WorkflowHub team resulted in the creation of automatic registration and metadata parsing functions for this community's Nextflow workflows\footnote{\url{https://elixiruknode.org/news/2024/workflowhub-nf-core-workflow-accessibility/}}. These workflows can now be found in WorkflowHub\footnote{\url{https://workflowhub.eu/workflows?filter[project]=15}}. 

\subsubsection{\textbf{Individual workflow developers}}

WorkflowHub supports large international science missions and consortia. However, the registry welcomes and encourages contributions from any workflow developer, regardless of their research domain, application areas, or the size of their research group. In fact, the largest Space on WorkflowHub is currently Independent Teams, which covers 287 people, 185 Teams, 180 Organisations and 279 workflows.

\section{Discussion}\label{sec:discussion}

We have presented WorkflowHub, a registry that enriches the scientific workflows ecosystem by being a hub for discovery and sharing of workflows from across multiple languages, communities, consortia, and scientific domains. The registry connects this community of users and contributors to workflow development, support, and scholarly services that support the requirement for workflows to be shared and credited, as well as the various requirements developers encounter during the workflow life cycle (see \textbf{Figure 1}). Workflows continually evolve to keep pace with changing research questions, data types and practices. Similarly, it is intended for WorkflowHub to evolve in lock step with the changing requirements of both the developers and users of computational workflows. The roadmap for WorkflowHub over the next few years can be broadly broken down into four aspects: improving the support and resources available to users of the registry, onboarding new communities and domains, contributing thought leadership for workflow best practices that directly impact aspects such as workflow visibility and quality, and aligning to FAIR principles for computational workflows.

\subsection{\textbf{Improving support}}

As WorkflowHub and its integrated services intend to support the workflow life cycle, and not simply the registration and publication of workflows, the registry will aim to improve processes that align to supporting this life cycle. This includes improving search and the inbuilt wizards that assist with onboarding and resource registration, but also improving overall ease-of-use and guidance. Additions and improvements for the UI will continue, adding those features that are required by end users. Particular attention will be paid to tracking workflow cloning, as well as supporting workflow collections, sub-workflows, and nested workflows. 

Continued work on the onboarding guidelines created for WorkflowHub will improve the quality of workflow organisation for contributing groups (i.e. big projects and consortia), in part by including guides that cover best practice instructions for the structure and content of workflow repositories as well as how to make the most impact with LifeMonitor, but also by contributing to training. To date, WorkflowHub has been included in existing training, such as the GTN Smörgåsbords\footnote{\url{https://training.galaxyproject.org/}} and workflow registration workshops \cite{gustafsson_2023_7787488}. As a next step, this effort should be extended to include WorkflowHub lessons in Software Carpentry developed by the Data Science training programme for Health and Biosciences (Ed-DASH\footnote{\url{https://edcarp.github.io/Ed-DaSH/}}) for Nextflow\footnote{\url{https://carpentries-incubator.github.io/workflows-nextflow/}} and Snakemake\footnote{\url{https://carpentries-incubator.github.io/snakemake-novice-bioinformatics/}}. 

As the number of contributors and workflow assets grow, it will be critical to also explore the use of automated mechanisms that allow scaling of support. This could include streamlined metadata annotation approaches that use Large Language Models (LLMs) to populate metadata fields for review by a workflow creator, automatic reporting of issues to workflow creators and their Teams, or integration with additional added value services like APICURON \cite{hatos_apicuron_2021}. 

Reflecting their importance to the operation of WorkflowHub, integrations with platforms and services that support the workflow life cycle will be added, improved, and updated. The RO-Crate format used by WorkflowHub will be central to this effort. Planned updates include the ability to natively configure automated synchronisation between Git repositories and WorkflowHub. Setting this up to match the expectations of the community of WorkflowHub users will be essential, and will depend on registry and community co-development, in particular for those case where work involves the repositories used by WfMS communities. In addition, the ability to “launch” workflows on additional instances that implement TRS will be explored (e.g. Seqera Platform\footnote{\url{https://seqera.io/platform/}}), and benchmarking service integrations for workflow entries will be added through collaboration with OpenEBench\footnote{\url{https://openebench.bsc.es/}} \cite{capella-gutierrez_lessons_2017}.

\subsection{\textbf{Onboarding}}

WorkflowHub has already supported the onboarding of WfMS communities, including Galaxy's IWC and Nextflow's nf-core. A similar integration is currently being finalised for the GTN. These community-centric engagements have resulted in the addition of hundreds of workflows to the registry and, given that WorkflowHub is now integrated with the source repositories where new workflows are developed, this number will only grow. Identifying and engaging workflow communities to onboard, in particular those that converge on specific WfMS, will be critical to increasing the number of workflows in the registry and the visibility of the workflows ecosystem overall, as well as to ensure that further requirements for WorkflowHub are being collected from a diverse set of stakeholders. Support provided to new communities may entail 1) tailored integrations with community repositories, facilitating semi-automated ingestion of workflows to WorkflowHub and synchronised registration of new workflow releases, and 2) supporting WfMS communities to make the best use of WorkflowHub, including how to implement high quality workflow repository structures and effectively adopt both standards (e.g. RO-Crate, Abstract CWL, TRS) and services (e.g. LifeMonitor). This level of support is now being explored for Snakemake workflows, and will involve close collaboration with the Snakemake community.

\subsection{\textbf{Workflow visibility and recognition}}

WorkflowHub already has the capacity to import citation, author and contributor credit metadata from CITATION.cff files. In future, this could be extended to other popular standards, including codemeta.json \cite{jones_codemeta_2024}. Through DataCite, WorkflowHub has the means to mint DOIs, and incorporate workflows with DOIs into OpenAIRE\footnote{\url{https://www.openaire.eu/}}. WorkflowHub has thus actively worked to increase the visibility of workflows in a standardised and streamlined fashion. This approach is already bearing fruit, with examples of WorkflowHub formatted citations appearing in the published literature \cite{hall_pangenome_2024,roach_hecatomb_2024}. In addition, there are now examples of journals making computational workflows the focus of published works. A critical outcome here is to ensure that WorkflowHub becomes a recommended registry for journals and publishers. This includes providing workflow creators and users with a set of best practice recommendations for how to properly document and ultimately cite workflows in published research. A forum has already been established with multiple publishers and journals, and these continuing conversations will aim to address the challenges (i.e. citation formats, complexity of workflow citations, impact on publishing system, recommended practices for peer review of workflows) and opportunities (i.e. developer recognition, tracking, reproducibility) presented by formal citation of computational workflows.

\subsection{\textbf{Hand-holding for FAIR principles}}

As frequently alluded to in other sections, WorkflowHub is strongly connected to the FAIR principles at numerous levels. One of these levels that is of particular importance for users is the way that it “holds users’ hands” during the process of sharing workflows and related research artefacts. The FAIR principles are only guidelines, and even the most well-intentioned attempts to follow them can go awry in unexpected ways \cite{wilkinson_f_2022}. WorkflowHub aims to help the user to follow best practices, including following the FAIR principles, by making them convenient, but not imposing them as requirements.

For example, WorkflowHub helps users make their workflows Findable – easy to find for both humans and machines – by assigning globally unique and persistent identifiers for the workflows and their different versions (i.e. WorkflowHub identifiers and DOIs). WorkflowHub also guides users to describe their workflows with rich metadata, including the identifier for the workflow, and these metadata are automatically exposed for indexing through WorkflowHub’s use of Bioschemas.

WorkflowHub similarly helps users make their workflows Accessible – available to humans and machines over open protocols that provide optional access control – by providing multiple APIs over HTTPS. These APIs include the JSON-based FAIRDOM-SEEK API\footnote{\url{https://workflowhub.eu/api}}, an RO-Crate Submission API\footnote{\url{https://about.workflowhub.eu/developer/ro-crate-api/}}, and the TRS API. Workflow DOIs can also be minted, which means the workflow and its metadata will be accessible even if the workflow itself is no longer available or if the workflow itself cannot be shared openly.

Workflows, after being found and accessed, should ideally be Interoperable – able to be used by humans and machines as part of a wider computational ecosystem. Interoperability often requires a lot of “plumbing” that WorkflowHub provides for users automatically through the use of open-source standards (i.e. RO-Crate as the primary data exchange format) and domain- and tool-specific integrations (i.e. Galaxy IWC and Nextflow nf-core). By guiding users through the process of inputting metadata, WorkflowHub reduces complexity and tedium, making it significantly easier to create interoperable workflows.

Finally, WorkflowHub helps users ensure their workflows are Reusable – allowed to be used in part or in entirety by other humans and machines. In particular, it does this by providing opportunities to specify a clear and accessible license, qualified references to other software, and detailed provenance. WorkflowHub also collaborates closely with prominent communities in the computational workflows space (see Use Cases section) so that the registry can accommodate and incorporate domain-relevant community standards.

\section{Methods}\label{sec:methods}

\subsection{\textbf{Governance}}

WorkflowHub is an ELIXIR service supported by the UK and Belgium ELIXIR Nodes\footnote{\url{https://elixir-europe.org/}} \cite{elixir_elixir_2024} as well as Australian BioCommons\footnote{\url{https://www.biocommons.org.au/}} \cite{francis_australian_2024}. The registry is part of both the ELIXIR Tools Platform\footnote{\url{https://elixir-europe.org/platforms/tools}} and Research Software Ecosystem\footnote{\url{https://research-software-ecosystem.github.io/}}, and forms part of both EuroScienceGateway\footnote{\url{https://esciencelab.org.uk/projects/eurosciencegateway/}} and DARE-UK TRE-FX\footnote{\url{https://esciencelab.org.uk/projects/tre-fx/}}. Governance is coordinated and managed by the WorkflowHub Club\footnote{\url{https://about.workflowhub.eu/project/community/}}: an inclusive community that meets biweekly online and consists of workflow developers/creators and workflow users, as well as WorkflowHub developers and product owners. The minutes for these meetings are open\footnote{\url{https://docs.google.com/document/d/1U2KAlbKviCu-fCX-znncKIBUIUUOeEnuRGdAg-fNd4Q/edit}} and a GitHub organisation is used to manage documentation\footnote{\url{https://github.com/workflowhub-eu/about}}. Club members include representatives from ELIXIR, WCI, Australian BioCommons, and more. Over 60 people are listed as contributors\footnote{\url{https://about.workflowhub.eu/project/acknowledgements/\#workflowhub-club}}, and any new contributors are invited to join. WorkflowHub has secured funding for sustainability through Horizon Europe projects, ELIXIR Europe and national UK funds. 

\subsection{\textbf{Technical infrastructure}}

WorkflowHub is developed openly, and largely virtually, using open software development practices, hackathons, and virtual communication channels. It has both a roadmap\footnote{\url{https://about.workflowhub.eu/project/roadmap/}} and regular release cycle (i.e SEEK release cycle\footnote{\url{https://github.com/seek4science/seek/}}. WorkflowHub requests for Team registration are supported for the American and Asia Pacific time zones by Oak Ridge National Laboratory\footnote{\url{https://www.ornl.gov/}} and Australian BioCommons, respectively. WorkflowHub is built on the FAIRDOM-SEEK software \cite{wolstencroft_fairdomhub_2017}. FAIRDOM-SEEK was originally developed as a data-management platform for the systems biology community, but has been generalised over time to support a wide variety of use cases, and now has numerous deployments across the world supporting many different communities. Development of FAIRDOM-SEEK is a collaborative activity with contributors from institutions in the UK, Germany, Belgium, Sweden and elsewhere. WorkflowHub is currently hosted on the University of Manchester's Research IT cloud. WorkflowHub makes use of Git to store workflows as repositories. This enables the addition, modification and deletion of files for workflows that are uploaded directly to the registry by a user, as well as the ability to freeze “snapshots” of specific workflow versions. If a workflow has been added either via Git import or through submission of an RO-Crate, it then needs to be explicitly versioned and uploaded again. Curation of registered workflows is the responsibility of the workflow creators and / or submitters.

\section{Data Availability}\label{sec:data-availability}

WorkflowHub will ensure data and metadata availability and access up to 2027, with plans to extend this availability further. Further availability of data and metadata from WorkflowHub falls under two categories. For workflows with minted DOIs, the archiving of data and metadata is managed via DataCite. For all workflows, and associated digital assets, WorkflowHub has an End-of-Life policy. The policy states that \textit{“If and when the WorkflowHub reaches its end of service after that (i.e. 2027), the published contributions and metadata will be archived as RO-Crates and made available through a public repository, such as Zenodo, Figshare or another appropriate resource at that time. DOI registrations will in this case be updated to link to the archived deposits.”}\footnote{\url{https://about.workflowhub.eu/project/\#retention-and-end-of-life-policy}}. A knowledge graph of registered Workflow RO-Crates as of 2024-08 is published on Zenodo \cite{hambley_workflowhub_2024}.

\section{Code Availability}\label{sec:code-availability}

WorkflowHub code is available as part of the FAIRDOM-SEEK project\footnote{\url{https://github.com/seek4science/seek}} \cite{owen_seek4scienceseek_2024}.

\backmatter

\bmhead{Acknowledgements}

workflowhub.eu was founded as part of EOSC-Life (WP2 Tools Collaboratory), funded by the European Union’s Horizon 2020 programme under grant agreement H2020-INFRAEOSC-2018-2 824087. This work was further supported by funding from European Commission’s Horizon Europe programme and UK Research and Innovation (UKRI) under the UK government’s Horizon Europe funding guarantee: EuroScienceGateway (HORIZON-INFRA-2021-EOSC-01-04 101057388, UKRI 10038963), BY-COVID (HORIZON-INFRA-2021-EMERGENCY-01 101046203), FAIR-IMPACT (HORIZON-INFRA-2021-EOSC-01-05 101057344, UKRI 10038992), BioDT (HORIZON-INFRA-2021-TECH-01-01 101057437, UKRI 10038930), PREP-IBISBA (H2020-INFRADEV-2019-2 871118), AgroServ (HORIZON-INFRA-2021-SERV-01-02 101058020, UKRI 10038927), BIOINDUSTRY 4.0 (HORIZON-INFRA-2022-TECH-01 101094287, UKRI 10048146). This work is supported by Australian BioCommons which is enabled by NCRIS via Bioplatforms Australia funding. This work is also funded by the Sardinian Regional Government through the XData Project, and by the LIFEMAP project (Italian Ministry of Health, POS T3). The work supported by the ELIXIR-DE node was supported by the German Federal Ministry of Education and Research BMBF grant 031 A538A de.NBI-RBC and the Ministry of Science, Research and the Arts Baden-Württemberg (MWK) within the framework of LIBIS/de.NBI Freiburg. This research used resources of the Oak Ridge Leadership Computing Facility at the Oak Ridge National Laboratory, which is supported by the Office of Science of the U.S. Department of Energy under Contract No. DE-AC05-00OR22725. The authors are grateful to Sehrish Kanwal of the University of Melbourne for feedback and suggestions about the draft. 

\bmhead{Author Contributions}
Here, we follow the Contributor Role Taxonomy (CRediT)\footnote{\url{https://credit.niso.org}}. OJRG contributed through: Writing -- original draft, Writing -- review \& editing, Project administration, Data curation, and Supervision. SRW contributed through: Writing -- original draft, Writing -- review \& editing, and Project administration. FB contributed through: Conceptualization, Data curation, Investigation, Project administration, Software, Methodology, and Resources. SSR contributed through: Conceptualization, Funding acquisition, Investigation, Supervision, and Writing -- review \& editing. SO contributed through: Conceptualization, Investigation, Project administration, Software, and Methodology. NJ contributed through: Data curation and Writing -- review \& editing. FC contributed through: Conceptualization, Funding acquisition, Methodology, Project administration, Resources, and Supervision. CG contributed through: Conceptualization, Funding acquisition, Methodology, Project administration, Resources, Supervision, Visualization, Writing -- original draft, and Writing -- review \& editing. All other authors contributed to the manuscript through: Writing -- review \& editing.

\bmhead{Competing Interests}

The authors declare no competing interests.

\bibliography{main}

\end{document}